\documentclass[12pt,preprint]{aastex}

\newcommand{\asca}{{\it ASCA}}
\newcommand{\ASCA}{{\it ASCA}}

\newcommand{\xmm}{{\it XMM-Newton}}
\newcommand{\XMM}{{\it XMM-Newton}}

\newcommand{\chandra}{{\it Chandra}}
\newcommand{\Chandra}{{\it Chandra}}

\renewcommand{\ion}[2]{\ensuremath{{\mathrm {#1}}\,{\textsc {#2}}}}

\shorttitle{Chandra HETGS observation of NLS1 Ark~564}
\shortauthors{C. Matsumoto et al.}

\begin{document}

\title{A Chandra HETGS observation of the Narrow-line Seyfert 1 galaxy 
Ark~564}

\author{Chiho Matsumoto and Karen M. Leighly}
\affil{Department of Physics and Astronomy, The University of Oklahoma,
    440 W. Brooks St., Norman, OK 73019}
\email{chiho,leighly@nhn.ou.edu}
\and
\author{Herman L. Marshall}
\affil{MIT Center for Space Research, 77 Massachusetts Ave., Cambridge,
 MA 02139}

\received{07/11/2002}
\revised{08/2?/2002}
\accepted{}

\begin{abstract}
We present results from a 50~ks observation of the  
narrow-line Seyfert 1 galaxy Ark~564 with 
the \chandra\ HETGS.
The spectra above 2~keV are modeled by a power-law
with a photon-index of $2.56\pm0.06$.
We confirm the presence of the soft excess below about
1.5~keV.
If we fit the excess with blackbody model,
the best-fit temperature is 0.124~keV.
Ark~564 has been reported to show a peculiar emission line-like
feature at 1~keV in various observations using lower 
resolution detectors, and the \chandra\  grating spectroscopy
rules out an
origin of blends of several narrow emission lines. 
We detect an edge-like feature at 0.712~keV in the 
source rest frame. 
The preferred interpretation of this feature is combination of 
the O~{\sc vii} K-edge and
a number of  L-absorption lines from slightly ionized iron,
which suggests a warm absorber with ionization parameter $\xi\sim1$ 
and $N${\sc h}$~\sim10^{21}$~cm$^{-2}$\@.
These properties are roughly consistent with those of
the UV absorber.
We also detect narrow absorption lines of O~{\sc vii}, 
O~{\sc viii}, Ne~{\sc ix}, Ne~{\sc x}, and Mg~{\sc xi}
at the systemic velocity. From these lines, a second
warm absorber having  $\log\xi\sim$ 2 and 
$N${\sc h}$~\sim10^{21}$~cm$^{-2}$
is required.

\end{abstract}


\section{Introduction}

Narrow-line Seyfert~1 galaxies (NLS1s) belong to a ubiquitous 
class of X-ray luminous AGNs whose extreme spectral and 
variability properties have
been the subject of intensive study by virtually every X-ray satellite
during the past decade.  Defined by their optical emission-line
properties (FWHM H$\beta < 2000$~km~s$^{-1}$ and
[\ion{O}{iii}] $\lambda$5007/H$\beta < 3$) 
NLS1s are different from
type~2 Seyfert galaxies, which generally have 
[\ion{O}{iii}]/H$\beta > 3$ (Osterbrock \& Pogge 1985; Goodrich 1989).  
NLS1s usually have strong permitted lines of \ion{Fe}{ii},
resembling the well-known prototype of their class, I~Zw~1.

The extreme X-ray properties of Narrow-line Seyfert 1 galaxies are now
well estabelished. They frequently show a strong soft excess component,
their hard X-ray spectrum tends to be steeper than in similar
broad-line Seyfert 1 galaxies, and they show enhanced X-ray
variability (e.g., Leighly 1999a,b and references therein).  
The most promising explanation
for this behavior is that NLS1s have a higher mass accretion rate
with respect to the Eddington value than ordinary Seyfert galaxies with 
broad optical lines \citep{laor00}.
This result is potentially very
important: since AGN are believed in general to be powered by
accretion, study of objects with the highest accretion rate may help
us understand AGN accretion in general.

Ark~564 (z$=$0.02468$\pm$0.00007; Huchra, Vogeley, \& Geller 1999) 
is an interesting object from the point 
of view that shows a peculiar emission line-like 
feature near 1~keV in low resolution spectra. 
This emission line-like feature has been reported 
from various observations performed by {\it ROSAT}, 
\asca\  and {\it BeppoSAX}
\citep {bra94, lei99b, tur99, com01}. 
Turner et al. (1999) modeled the feature with a strong 
(equivalent width: EW$\sim$70~eV)
and relatively 
broad ($\sigma$=0.16~keV) line at 0.99$^{+0,02}_{-0.04}$~keV,
assuming that spectral components are Galactic absorption, 
a power-law and a Gaussian line 
in the energy band of 0.6--5.0 and 7.5--10~keV.
Using the same data and the different continuum model
(Galactic absorbed a power-law plus a black body), 
Leighly (1999b) obtained weaker (EW$=30$~eV) line 
at 1.033$^{+0.024}_{-0.031}$~keV with $\sigma$ of 0.1~eV.
\citet {com01} also reported that
the model dependent EW is as low as 20--50~eV 
with the same model as Leighly's from independent analyses.
In the {\it BeppoSAX} observation,
a line-like feature around 1~keV is clearly visible \citep {com01}.
Still the origin of this feature was far from being unambiguously 
identified.

In order to identify this feature, we carried out an 
observation with the \chandra\ High Energy Transmission Grating 
Spectrometer (HETGS).
Observed with higher energy resolution, some
possible origins for this feature, such as blends of narrow emission lines, 
can be investigated.
If we can identify this feature, it may enable
us to investigate the physical conditions such as ionization state
of the surrounding matter of the AGN.
Ark~564  is one of the brightest narrow-line Seyfert 1 
galaxies in the hard X-ray band and the 2--10~keV flux
is a few times 10$^{-11}~{\rm erg~cm^{-2}~s^{-1}}$, so a
grating observation using \chandra is clearly feasible.

In this paper, we will report the fine spectroscopy of Ark~564.
Section 2 describes the details of the observations and data reduction.
In \S3, the analysis of power spectral density is performed.
\S4 contains the spectral analysis. 
We detected many absorption lines, but 
no prominent narrow emission line around 1~keV is detected.
In \S5, we discuss the inferred photoionization model and
the location of warm absorber. 
We also find that the 1~keV emission line feature 
appears to be an artifact of the warm absorber.
Finally, we summarize the results in \S6.

\section{Observations}

We carried out an observation of Ark~564
with the \chandra\ 
HETGS 
on 2000 June 17 06:50--21:05.
The detector was the Advanced CCD Imaging Spectrometer
(ACIS). 
Our observation was 50~ks in duration and continuous.
We analyzed the most recent standard processed level 2 data 
(revision 3)
using the CIAO version 2.2.0, CALDB 
version 2.9, and HEAsoft version 5.1.

The peak of the zeroth order image is located
at 
R.A.$={\rm 22^h 42^m 39.33^s}$ and 
Decl.$=29^\circ 43' 31\rlap{$''$}.6$
(equinox 2000.0)
in good agreement with the optical  position of 
R.A.$={\rm 22^h 42^m 39.345^s}$,
Decl.$=29^\circ 43'  31\rlap{$''$}.31$ (Clements 1981)
with the \chandra\  spatial resolution (0.49~arcsec/pixel).

Since the HETGS response matrices are subject to
the extension of the source,  
we checked whether or not the source is point-like. 
At first, we looked at the zeroth order image. 
We found the image is circular in shape, thus asymmetry 
possibly associated with extended emission is ruled out.
Next, we tried to investigate point spread function, but
it is difficult with the zeroth order image,
since the central and adjacent pixels suffer significant pile-up.
Then, we examined the first order data in spatial direction and 
found that 67~\% of photons lie
within $\pm$1 pixel and 87~\% lie within $\pm$2 pixels.
The encircled power radii of 50 and 80~\% are 0.418 and 
0.685~arcsec (0.85 and 1.4 in the unit of pixel), 
respectively\footnote{http://cxc.harvard.edu/cal/Hrma/hrma/misc/oac/psf2/}.
Therefore, the core of the X-ray emission is not extended,
and we used the response matrices for a point source
in the spectral analyses.

\begin{figure}
\plotone{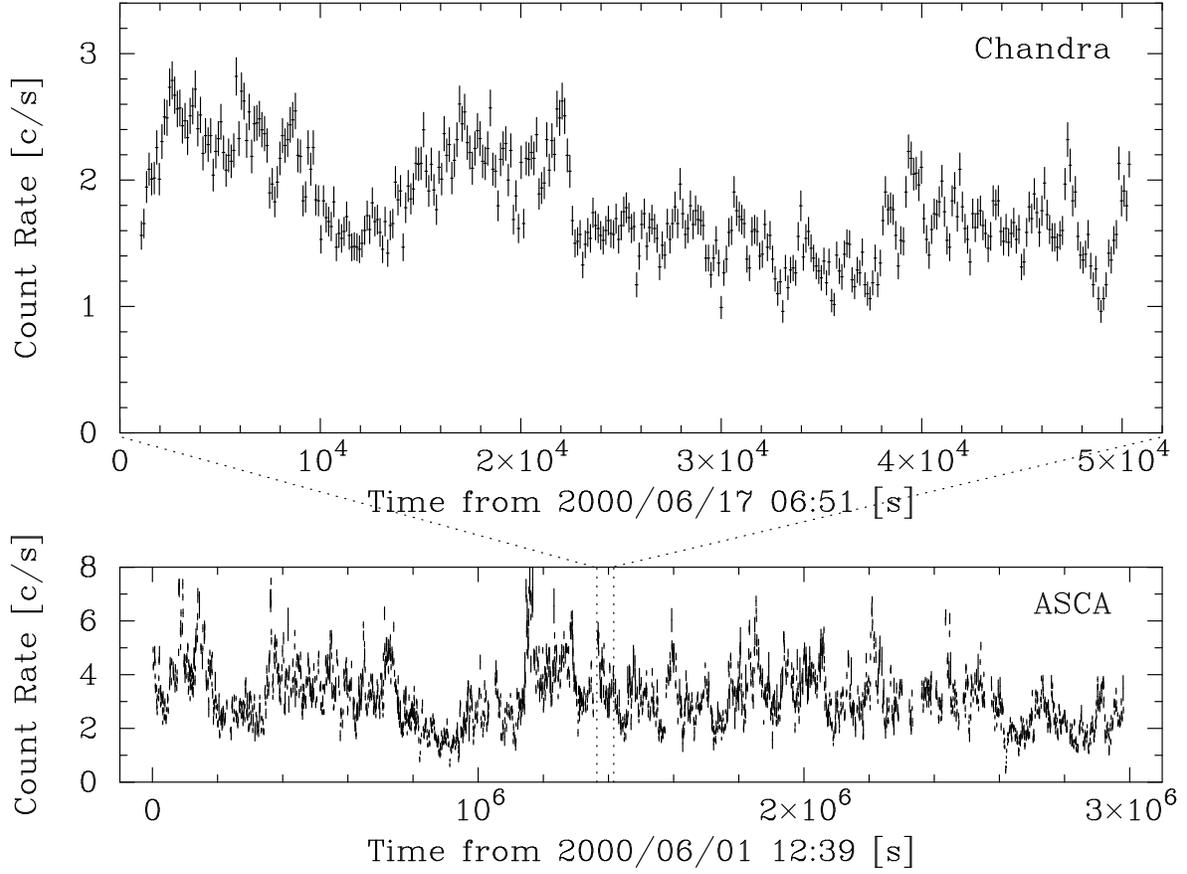}
\caption[h]{
Upper: The lightcurve from the \chandra\ HETGS binned at 128~s. 
The positive and negative first order data from both the HEG and the MEG 
are combined. 
Lower: The lightcurve from the \asca\  SIS detectors. 
The period of our \chandra\ observation is marked by dotted lines.
The energy range of HETGS and SIS are 0.4--10 and 0.5--10~keV, respectively. 
}
\label{fig:lc}
\end{figure}

Fig.~\ref{fig:lc} shows the lightcurve from the \chandra\  HETGS.
We use \chandra\  data only from the MEG and HEG $\pm1$ orders.
The count rate is somewhat higher during 
the first half of the \chandra\  observation.
Simultaneously, there was a multi-wavelength campaign on this source 
using {\it ASCA}, {\it RXTE}, {\it HST}, and ground base telescopes
(e.g., Turner et al. 2001 and Edelson et al. 2002).
We downloaded the screened \asca\  data from GSFC rev2 archive
and created the lightcurve from June 1 to July 7.
To evaluate the 2--10~keV flux state during the \chandra\  observation, 
we compare the count rate and variability from \asca\ in the \chandra\ 
observation with those in the whole \asca\  observation.
We summarized the average count rate and the fractional 
RMS variability in Table~\ref{tab:var}. 
During the \chandra\  observation, 
the source was slightly brighter than average by about 20~\% 
and variability is typical, compared with the \ASCA\ observation.

\section{Power Spectral Density}

An identifying feature of active galaxies is their X-ray variability.
Their variability is not periodic but rather a featureless power law
on time scales of weeks to days (e.g., Lawrence \& Papadakis 1993).
This is unfortunate, because no time scales that might correspond to
physical size scales such as the size of the emission region, can be
found.

As well as fine spectroscopy, an advantage of the \Chandra\
observatory is the continuous data sampling, which allows us to
compute power spectral density (PSD) directly in the frequency range
around $2\times10^{-4}$ Hz. In principle, because of the continuous
data sampling and fairly large effective area, power spectral density
analysis should allow us to observe the power spectrum on short enough
time scales to look for breaks that may be indicative of a physical
size scale.

\begin{figure}
\plotone{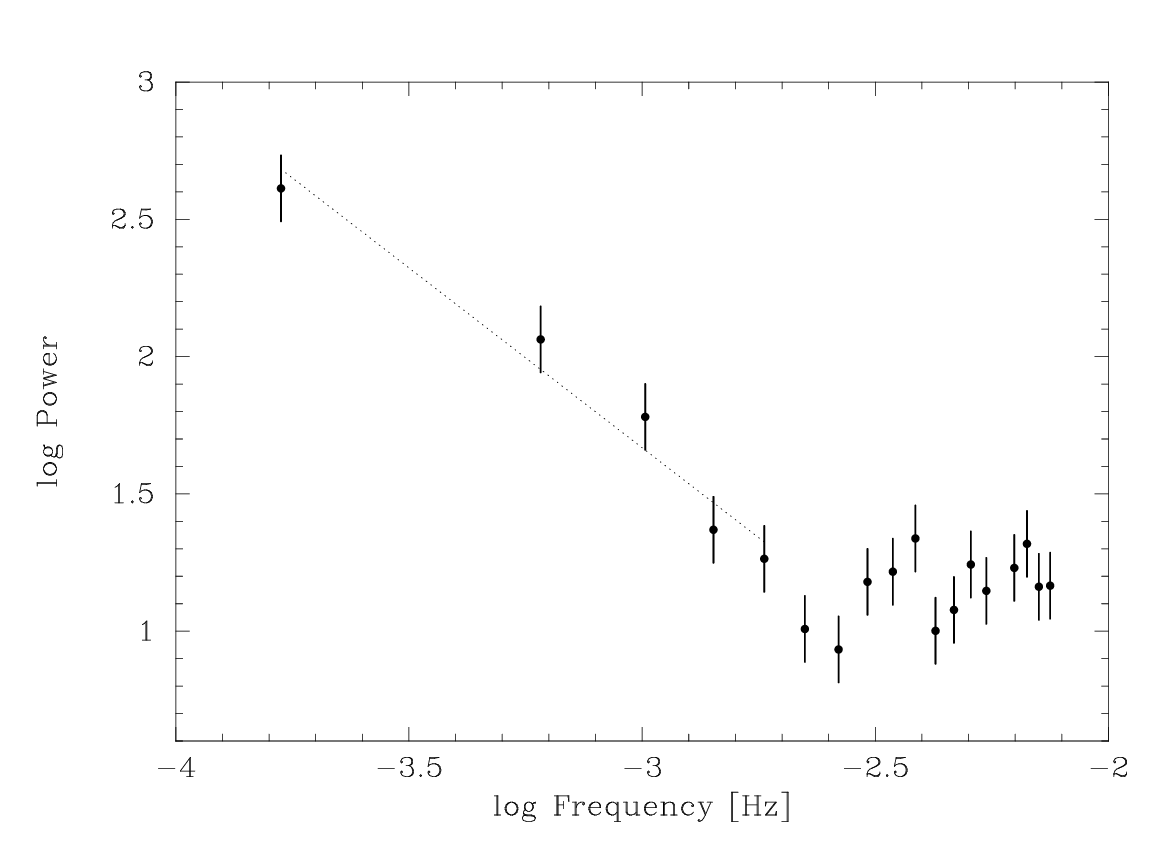}
\caption{Power density spectrum of Ark~564 during the \chandra\ observation.
We grouped the data and calculated the errors
in the manner described in \citet {papa93}.
The dotted line shows the best-fit power-low model in the frequency range of
$1.7\times10^{-4}$ to $2.0\times10^{-3}$~Hz. 
The power-law index is  $1.31\pm0.24$. 
}
\label{fig:pds}
\end{figure}

We created the lightcurve binned at 64~s and calculated the power
spectrum in the frequency range between $2.0\times10^{-5}$ and
$7.8\times10^{-3}$~Hz.  Fig.~\ref{fig:pds} shows the PSD of Ark~564
during the \chandra\ observation.  In order to obtain reasonable
signal to noise, we grouped in sets of 20 data points and averaged
their logarithm \citep {papa93}.  The background due to Poisson noise
is not subtracted and we find that it dominates the frequencies
above $10^{-2.7}(=2.0\times10^{-3}$)~Hz.  Below this frequency, the
fitted power-law index ($\alpha$ where PSD is proportional to
$\nu^{-\alpha}$) is $1.31\pm0.24$. 
This is rather typical for AGNs on short time scales \citep{law93}.

Using the 0.7--5~keV data from the \asca\ long-look observation,
\citet{papa01} detected a high frequency break at
$(2.3\pm0.6)\times10^{-3}$~Hz and a slope of 1.24$^{+0.03}_{-0.04}$
below the break frequency.  Although the break frequency cannot be
detected in our data due to the Poisson noise, the slope that we
measure is consistent with their result.

\section{Spectral Analyses}

We analyzed the MEG$\pm$1 and HEG$\pm$1 spectra.
The errors quoted in this section are 90~\% 
for a single parameter of interest ($\Delta\chi^2=2.71$). 
Calibration errors for absolute flux are not taken into account;
They are estimated to be less than 10~\% and 20~\% 
for the 1.5--6~keV band and the most of the other energy bands, 
respectively\footnote{e.g., http://space.mit.edu/CXC/calib/hetgcal.html}.
Calibration uncertainty for the relative flux between HEG and MEG
is accounted for by introducing a constant parameter into the model.

\subsection{Continuum} \label{sec:cont}

\begin{figure}
\plotone{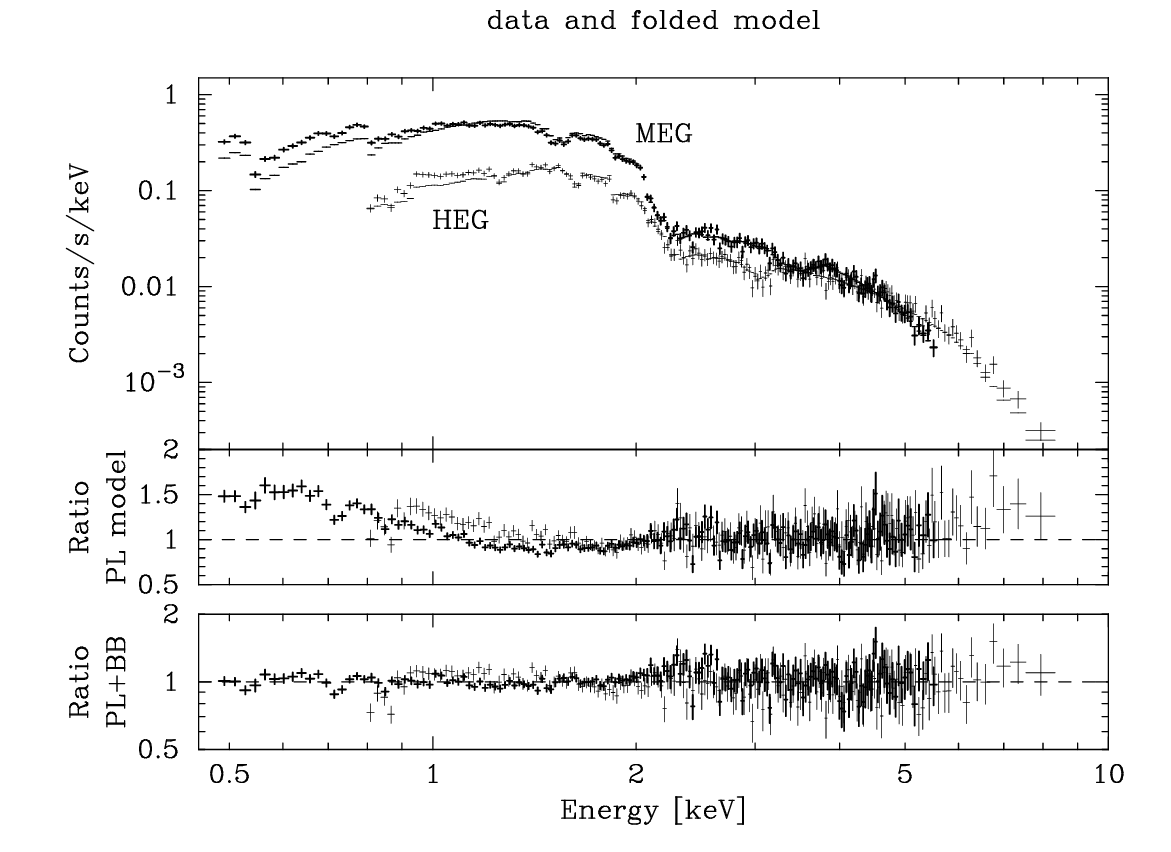}
\caption[h]{Top: The HETGS spectra and
the best-fit power-law model in the 2--5~keV band. 
The MEG$\pm$1 and HEG$\pm$1 are thick an2d thin lines, respectively.
Middle: The ratios of data to the power-law model 
extrapolated to the whole energy band. 
Bottom: The ratios to the power-law plus black body model.
All models include Galactic absorption.
}
\label{fig:pl}
\end{figure}

First, we tried to fit the data
in the whole energy band with a model consisting
of a power-law and Galactic absorption 
($N${\sc h} $=6.4\times10^{20}~{\rm cm}^{-2}$; 
Dickey \& Lockman, 1990\footnote{
http://heasarc.gsfc.nasa.gov/cgi-bin/Tools/w3nh/w3nh.pl
}). 
To look at the large scale behavior, 
we rebinned the spectra coarsely,
so that the energy resolution is similar to that of 
the {\it ASCA}  SIS,
and so that each energy bin contains more than 25~photons.
The residuals showed systematic humps with amplitude larger 
than the calibration errors in the softest and hardest energy bands.
First we fitted the data only in the 2--5~keV band 
and obtained a reasonable $\chi^2$ value of 96.1 for 110 
degrees of freedom (d.o.f.).
The photon-index and the unabsorbed 2--10~keV flux are 
$2.56\pm0.06$ and 
(2.42$\pm$0.06)$\times10^{-11}~{\rm erg~cm^{-2}~s^{-1}}$, 
respectively.  
Fig.~\ref{fig:pl} shows the data and the best-fit model 
mentioned above and extrapolated to the whole energy band. 
Excess emission below about 1.5~keV is seen clearly 
in the ratio plots.

We tried to model the soft excess component in three ways.
Throughout we fixed neutral absorption to the Galactic value.
First, we fitted data with a broken power-law and neutral 
absorption model. 
The fit gives a break energy of 1.45$\pm$0.04~keV 
and a photon-index below the break energy of 3.11$\pm$0.03.
The reduced $\chi^2$ value is 1.40 for 359 d.o.f.
Next we tried to fit with a two power-law model.
This fit gives worse $\chi^2_\nu$ value of 1.69 for the same 
number of d.o.f.
The crossing point of the two power-laws is 1.39~keV.
Finally we fitted the soft excess component with 
a black body model.
The fit gives the blackbody temperature of 0.124$\pm$0.003~keV 
in the source rest frame
and a better  $\chi^2_\nu$ value of 1.34 with the same 
number of d.o.f.
The fits are still not acceptable.
This could be due in part to the edge-like feature around 0.7~keV
that can be seen in Fig.~\ref{fig:pl}.

\subsection{Iron K-emission Line}

Although statistically limited, 
HETGS data is useful to investigate narrow iron K-emission lines.
We created  HEG $\pm$1 and  MEG $\pm$1 spectra
binned at the detector resolution (FWHM).
The spectra have 2--13~photons per energy bin around 
the iron K-energy band.
Since the data are in the Poisson regime, we performed the spectral 
fitting and error analysis using the $C$ statistic.
We fitted the data in the 6.0--7.0~keV energy band
with a power-law plus three Gaussian lines 
assuming all the lines are narrow and at the systemic velocity.
Due to poor statistics, we performed fits 
fixing the power-law index at several values.
With the photon-index of 2.56 
(the best-fit value from the 2--5~keV band fit 
in \S\ref{sec:cont}),
the fit gives upper-limits of the EWs of 
45~eV, 53~eV and 78~eV
for the neutral, He-like and H-like lines, respectively.
The upper-limits are weakly subject to the assumed 
continuum slope, but the difference is as small as 5~eV
even in a wide range of photon-index of 2.0--3.0.
Thus, we can conclude that 
no strong narrow iron K-emission lines are detected;
hence, the diskline feature with huge EW 
(653$\pm$85~eV for Ly$\alpha$) reported by \citet {tur01}
is not significantly contaminated by narrow lines.

\subsection{Fine Spectral Features in the Soft X-ray Region}

We investigated narrow features  in the soft X-ray region
using the HETGS high resolution spectra.
We binned the HETGS first order spectra according to the following criteria:
the energy bin widths are at least as large as the detector 
energy resolution,
and each energy bin contains at least 25~photons.
The latter criterion allows us to use $\chi^2$ fitting.

First we looked at the edge-like feature around 0.7~keV.
If we add an edge to the model, the fit is improved
significantly (99.9\% by F-test). 
The edge energy is  $0.711^{+0.004}_{-0.002}$~keV 
in the source rest frame and the optical depth is
$0.12^{+0.05}_{-0.04}$.
This energy is close to that of O~{\sc vii} K-edge (0.739~keV) 
but is inconsistent at the significance level of more than 99.9\%.

\begin{figure}
\plotone{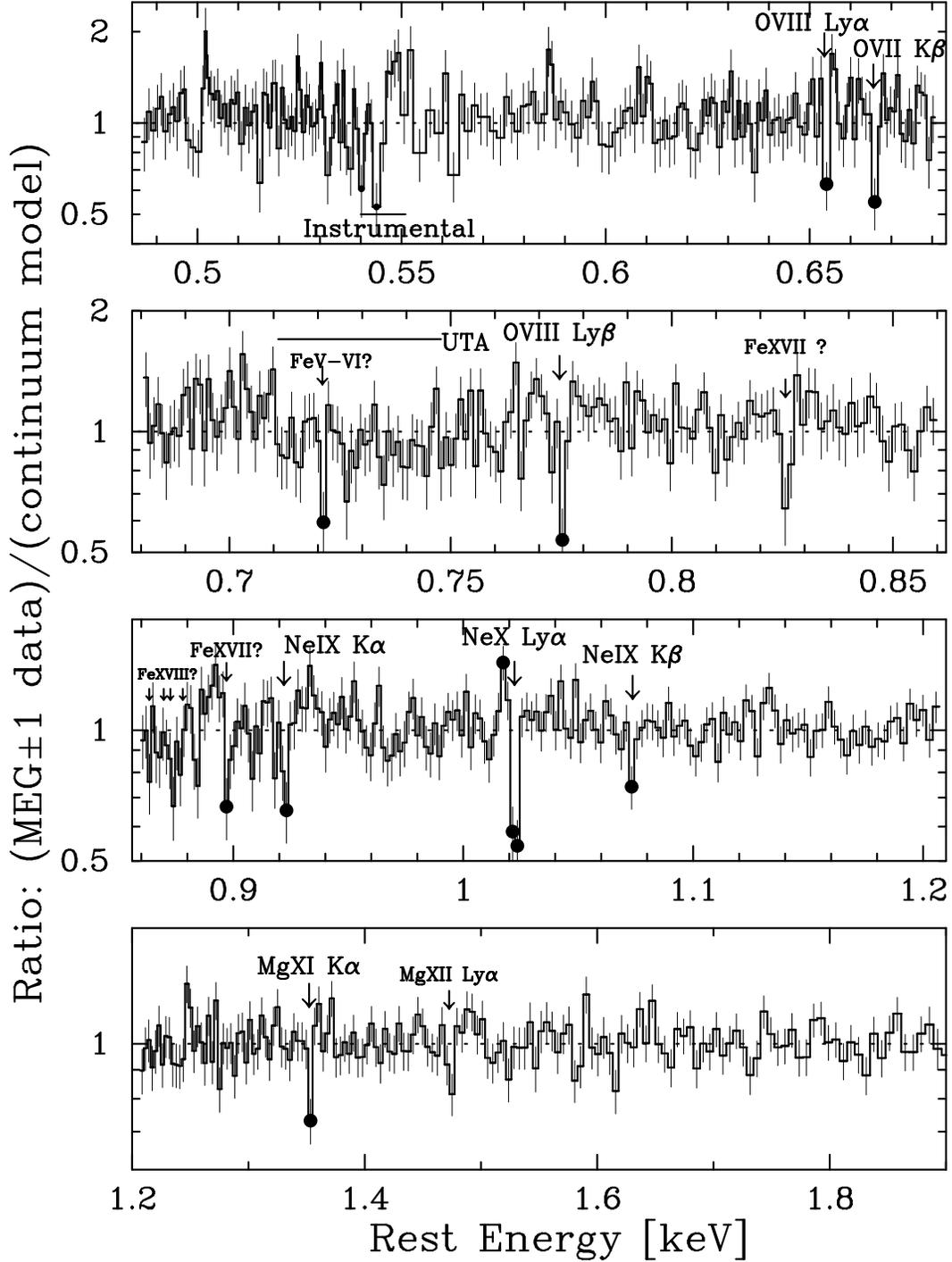}
\caption{The ratio of MEG$\pm$1 data to 
a continuum model.
Features detected at significance level of more than 3.0~$\sigma$
are marked by  $\bullet$.
A broad deficit also can be seen between 0.71--0.75~keV.
The labels with ``?'' denote suggested IDs.
}
\label{fig:fine}
\end{figure}

Next we searched for narrow features.
We estimate the continuum level by fitting the spectra
with a  power-law and Galactic absorption model
locally in each 0.15~keV band.
Fig.~\ref{fig:fine} shows the ratio of the MEG$\pm$1
data to the continuum model.
Most of narrow absorption features
(O~{\sc viii} Ly$\alpha$,
O~{\sc vii}~1s$^2$--1s3p,
Fe~{\sc xvii},
Ne~{\sc ix}~1s$^2$--1s3p
and Mg~{\sc xi}~1s$^2$--1s2p)
are detected at the energies consistent with the lines at the systemic redshift.
Some absorption lines of O~{\sc viii} Ly$\beta$, Ne~{\sc ix}~1s$^2$--1s2p, Ne~{\sc x}~Ly$\alpha$
are observed at slightly higher energies, suggesting 
blue-shift velocity is about 200~km~s$^{-1}$. 
However, 
taking into account both statistical error and 
HETGS capability\footnote{
According to The Chandra Proposers' Observatory Guide, 
systematic wavelength errors are at the 100~km~s$^{-1}$ level.},
this velocity-shift is rather marginal.
All the lines are unresolved (FWHM below $\sim$400~km s$^{-1}$).
The EWs are summarized in Table~\ref{tab:line}.
The errors of the EWs are calculated by scaling the line normalization.
Since these ions can make more absorption lines 
via other transitions,
we searched for all lines from these ions from the transition of 
$\alpha, \beta$ and $\gamma$ 
by fitting with additional absorption lines fixed at 
the expected energies.
The results are also listed in Table~\ref{tab:line}. 

There is a possibility that the line we identify as 
Ne~{\sc x}~Ly$\alpha$
is seriously contaminated by an  Fe~{\sc xvii} line.
The strongest predicted line of Fe~{\sc xvii} is at 0.826~eV and
it is marginally detected with EW of 0.7$\pm$0.4~eV.
Another strong line feature from Fe~{\sc xvii} at 0.726~eV
is marginally seen in Fig~\ref{fig:fine}.
These suggest that Fe~{\sc xvii} exists 
along our line of sight. 
We estimated the contribution to  
the Ne~{\sc x}~Ly$\alpha$ line 
to have EW of $\sim0.5(<1)$~eV
by scaling the EW of 
another Fe~{\sc xvii} absorption line at 1.011~keV 
that should be slightly weaker 
according to  the transition probability\footnote{
NIST atomic spectra databese;
http://physics.nist.gov/cgi\-bin/AtData/main\_asd}
than the one blended with Ne~{\sc x}~Ly$\alpha$\@.

\subsection{Edge-like Feature at 0.712~keV}

Edge-like features near 0.71~keV are frequently identified as
absorption associated with the O{\sc vii} K-shell.
The hypothesis that we observe  extremely 
redshifted \ion{O}{vii}  K-edge with a velocity 
of 10$^4$~km~s$^{-1}$ can be ruled out, because the detected \ion{O}{vii}
 absorption lines have no measurable velocity redshift 
(less than 122~km~s$^{-1}$).
Another possibility is that the feature is composed of a 
combination of 
absorption lines from the \ion{O}{vii}  Lyman transition, 
an \ion{O}{vii} K-edge
and neutral iron L-edges 
as in the Seyfert 1 galaxy MCG$-$6-30-15 \citep{lee01}. 
Their data show quite strong features from 
both the \ion{O}{vii} K-edge and K-$\gamma,\delta$ lines.
However, in our case, the \ion{O}{vii} K-$\gamma$ line is not detected
and its EW upper limit of 0.32~eV is 
one half of the best fit EW of \ion{O}{vii} K-$\beta$.
Thus the column density of \ion{O}{vii} is inferred not to be
high enough to produce significant lines of $\delta$ 
and higher.
Consistently, the $\tau_{\rm OVII}$ of $\sim$0.1 in Ark~564 
is much smaller 
than the $\tau_{\rm OVII}$ of $\sim$0.7 in MCG$-$6-30-15.
Thus, both the redshifted edge and the Lyman transition 
of the \ion{O}{vii} are unlikely to play a major role in  
this feature in Ark~564.

There is an alternative interpretation for the edge-like 
feature at 0.71~keV:
it is not an edge but rather the blue cut-off of the 
\ion{O}{viii} diskline emission \citep{braRay01}.
It should be noted that the iron K-emission diskline of Ark~564
observed by \asca\  \citep{tur01} was predominantly from H-like iron.
So, if this is the case, the innermost  part of the disk should
consist of two phases with quite high and moderate ionization parameter.
This model is difficult to investigate only with our \chandra\  data
and should be examined further using  \xmm\  RGS spectra.

With \xmm\ and \chandra, 
deep troughs are found in the soft X-ray spectra of 
several Seyfert galaxies.
They are inferred to be a number of 
iron L-absorption lines, the so-called Unresolved Transition Array 
(UTA) (e.g., Sako et al. 2001). 
If we parameterize the edge-like feature with a broad Gaussian,
the fit gives  central energy ($E_c$), velocity width ($\sigma$), 
and EW of 0.738$^{+0.06}_{-0.08}$~keV, 
11~eV, and $4$~eV, respectively.
If we add the \ion{O}{vii} K-edge 
at the energy fixed at the systemic velocity, 
the fit gives $\tau_{\rm OVII}$ of $0.07^{+0.04}_{-0.05}$
and the absorption line parameters of 
$E_c=0.730\pm0.06$~keV (16.98$\pm$0.14\AA), $\sigma=11\pm4$~eV, 
and EW$=4\pm2$~eV (93~m\AA). 
Since we detected the \ion{O}{vii} absorption lines, the latter model 
is more realistic.  
The central energy and the EW suggest that the major population of iron is \ion{Fe}{vii} 
and that the column density of iron 
is $\sim3\times10^{16}$~cm$^{-2}$ \citep{beh01}.
The presence of \ion{Fe}{v-vi} is also suggested from 
a significant absorption line at the energy of 
0.721~keV (17.20~\AA).
The \ion{Fe}{ix} ion, if the ionic fraction of it 
is similar to that of \ion{Fe}{v-vi}, 
should make detectable lines at 16.510~\AA\ (0.7510~keV)
and 16.775~\AA\ (0.7391~keV), and we do not detect them. 
Thus, we infer the ionization parameter $\xi$
\footnote{$\xi$ is defined as $L/nr^2$ [erg~cm~s$^{-1}$], where
$L$ is the hydrogen ionizing  luminosity, $n$ is 
the density, and $r$ 
is the separation between the absorber and the source of 
ionization radiation.}
of this absorber to be $\sim$1.
Assuming cosmic abundances, 
the column densities are estimated to be 
$N${\sc h}$\sim10^{21}$~cm$^{-2}$.
This photoionization model is roughly consitent with 
the ``lukewarm absorber'' 
(dimensionless ionization parameter 
$U$
\footnote{$U$ is defined as $\frac{Q}{4\pi r^2nc}$ where
$Q$ is the total number of hydrogen ionizing photons per second
 and $c$ is the speed of light. For the Ark~564 spectrum,
$U$ is smaller by about 2 orders of magnitude than $\xi$. }
$=0.032$ and $N${\sc h}$=1.62\times10^{21}$~cm$^{-2}$)
inferred from the absorption system in the UV band
\citep{cre02}.
Their prediction of $N_{\rm OVII}=1.5\times10^{17}$~cm$^{-2}$
is also consistent with 
the observed $\tau_{\rm OVII}$, which is equivalent to 
$N_{\rm OVII}\sim2.5^{+1.8}_{-1.9}\times10^{17}$~cm$^{-2}$.
Hence, we conclude that this feature is due to 
a warm absorber of $\xi\sim1$ and $N${\sc h}$\sim10^{21}$~cm$^{-2}$.

\section{Discussion}

\subsection{Physical Condition of the Higher-$\xi$
 Warm Absorber}

The low-$\xi$ warm absorber discussed in the previous section could 
be responsible for the narrow absorption lines of O~{\sc vii};
however, it is difficult to produce
the other higher ionization lines from O~{\sc viii}, 
Ne~{\sc ix}, Ne~{\sc x}, and Mg~{\sc xi} that we see.
This is because,
if the line-of-sight material is a single homogeneous gas,
there should be some observable lines 
from Ne~{\sc vi-vii} and Mg~{\sc vi-vii} that are not seen.
Therefore we infer that a second absorber with higher ionization
is present.
The suggested detection of the \ion{Fe}{v-vii} and \ion{Fe}{xvii}
lines and the lack of the \ion{Fe}{ix} lines 
also support the multi-phase/zone view.

An order of magnitude estimation of 
the column density ($N_{\rm ion}$)
and the velocity dispersion ($\sigma_{\rm ion}$) of the ion
for the higher-$\xi$ warm absorber
can be made using  curve of growth analysis.
The detection of a pair of absorption lines from the same ion
enables us to constrain $\sigma$ and $N$, because
the equivalent width $EW$ of an absorption line 
is a function of $N$, $\sigma$, and the oscillator strength
(Spitzer 1978; Kotani et al. 2000). 
It should be noted that the EWs of the 
K$\beta$ lines
are as large as the K$\alpha$ lines for 
O~{\sc viii} and Ne~{\sc ix} ions.
To explain the high value of the 
$EW_{\rm K\alpha}$/$EW_{\rm K\beta}$,
at least the K$\alpha$ lines of these ions
must be saturated.
This gives the upper and lower limits on $\sigma$ and  $N$,
respectively. 
The K-$\delta$ lines are not useful due to poor statistics. 
The inferred column densities are 
$N_{\rm O VII}\sim10^{17}~{\rm cm}^{-2}$, 
$N_{\rm O VIII}
\sim10^{18}~{\rm cm}^{-2}$, 
$N_{\rm Ne IX}\sim10^{17}~{\rm cm}^{-2}$,
$N_{\rm Ne X}\sim10^{17}~{\rm cm}^{-2}$, and 
$N_{\rm Mg XI}\sim10^{16.5}~{\rm cm}^{-2}$.
The velocity dispersion is $\sigma\sim$100~km~s$^{-1}$.
Using these column densities and assuming cosmic abundances, 
we infer that the photoionizd gas has 
the ionization parameter $\log\xi\sim2$
and the column density $N${\sc h}$\sim 10^{21}$~cm$^{-2}$.

\subsection{The Location of the Warm Absorbers in Ark~564}

The column density and ionization parameter of the low-$\xi$ warm
absorber seems to be consistent with that inferred from the high
resolution UV data (Crenshaw et al.\ 2002).  Therefore, it is
plausible that the low-$\xi$ X-ray warm absorber occurs in the same
gas that is responsible for the UV absorption.  

Crenshaw et al.\ (2002) note that the high resolution {\it HST} STIS
spectrum is unresolved in the spatial direction in both the continuum
and the emission lines, and 85\% of the Ly$\alpha$ flux comes from a
region smaller than $0\rlap{$''$}.2$.  From this they infer that
the narrow-line region in this object is more compact than 
95~pc.
Narrow-line region (NLR) sizes of $\sim 100\rm\,pc$ appear 
to be rather typical in Seyfert
1 galaxies (Schmitt \& Kinney 1996). Because the Ly$\alpha$ and
\ion{C}{iv} absorption lines are saturated, and are not filled in by
emission from the NLR, Crenshaw et al.\ infer that 
the absorber must fully cover the narrow-line region.  
They also note that
the inferred ionization parameter and density for the warm absorber
are such that the warm absorbing gas should emit optical lines.  
They
are forced to conclude that the global covering factor for the warm
absorber is 0.05 or less so that the strength of the predicted 
optical
emission lines are consistent with that observed.  
They infer that for
a distance to the warm absorber of more than 95~pc, the density
should be $\leq 10^3 \rm \, cm^{-3}$.

We note that if the NLR is as large as 95~pc, and the warm
absorber is present only toward our line of sight (i.e., the
absorption is not axisymmetric), the distance to the absorber 
may be as
small as 95~pc and still present a global covering fraction of
only 0.05. We note that despite the fact that UV absorption 
occurring along with X-ray absorption is found in about 50\% of 
active galaxies
(Crenshaw et al.\ 1999), an absorber presenting a covering factor of
0.05 in a random direction cannot be commonly seen 
in active galaxies,
simply because the small covering fraction means that 
the probability
of finding the warm absorber between the viewer and the nucleus is
low.

The likelihood of seeing such absorption may be increased if the
absorber is associated with a component that we know is already
present in AGN. Therefore, we suggest the possibility that the warm
absorber for Ark~564 may occur within the narrow-line region itself.
The absorption troughs may not be filled in by Ly$\alpha$ and
\ion{C}{iv} produced in the narrow-line region 
if those lines are not
strong in Ark~564.  As discussed by Crenshaw et al.\, Ark~564 is
heavily reddened.  If the dust responsible for the reddening is
located in the narrow-line region, that dust can reduce 
the resonance
line emission by absorption (e.g., Voit, Weymann \& Korista 1993; 
Hamann, Korista, \& Morris 1993; Ferguson et al.\ 1997). 
Other evidence discussed by Netzer \&
Laor (1993) points to the presence of dust in the NLR.  
The ionization parameter and density inferred for
the absorber by Crenshaw et al.\ are not terribly different than those
inferred for the narrow-line region (e.g., Kraemer et al.\ 1994).

The high-$\xi$ warm absorber may be coincident with the low-$\xi$ 
warm absorber. If so, a lower density by 2 orders of magnitude 
would be inferred.  Circumstantial evidence for this scenario is 
the fact the velocity offsets of the UV and X-ray absorbers are 
consistent with one another, as far as we can tell.  
It should be noted that 
the outflow velocity of 106--197~km~s$^{-1}$ revealed 
by the {\it HST} STIS echelle \citep{cre02} is difficult to 
be detected by the HETGS because of the relatively poorer 
data quality of {\it Chandra} compared with {\it HST}.

However, it is not necessary for the high-$\xi$ warm absorber to be
coincident with the low-$\xi$ one.  Close examination of the regions
of the spectrum right around the absorption lines reveals suggestions
of emission line wings tailing toward higher and lower energies
(e.g., the higher and lower energy side of the O~{\sc viii}
Ly$\alpha$ and Ne~{\sc x} Ly$\alpha$ lines, respectively),
although most of these features are not statistically significant.
These structures resemble those in the UV in which the center of the
emission lines has been removed by the nearly saturated broad
absorption lines such that only the high and low energy wings of the
line remains (Crenshaw et al.\ 2002).  The warm absorber shows us
ionized gas along our line of sight and the same photoionization may
also power emission lines in the gas predominately out of our line to
the central source if the covering fraction is large enough.

\subsection{Comparison of the Warm Absorbers with those in other NLS1s}

\Chandra\  grating observations have been reported from three NLS1s.
In NGC~4051, \citet{collinge01} found many absorption lines
from highly ionized ions such as oxygen, neon, silicon  and sulphur. 
They resolved two distinct absorption systems with 
high outflow velocity ($2340\pm130$~km s$^{-1}$) and 
low outflow velocity ($600\pm130$~km s$^{-1}$).
On the other hand, no absorption lines were detected in 
the spectrum of Ton~S180 \citep{tur01} and Mrk~478 \citep{mar03}.
The properties of the absorber in Ark~564 are different from both of them.

Although it is properly classified as a Seyfert 1 galaxy
and  has relatively narrow optical lines,
the features reported from MCG$-$6-30-15 by \citet {lee01}
are similar to what  we found in Ark~564. 
Specifically, they also detected narrow absorption lines 
from He-like and H-like oxygen
without detectable velocity shifts.
Moreover, 
in the O~{\sc vii} K-edge range, the deficit starts at 
an energy lower by about 30~eV than the energy of 
O~{\sc vii} K-edge. 
Absorption lines possibly from ionized iron are also detected.
All of these features are common between MCG$-$6-30-15 and Ark~564.

To date, results from the \XMM\ RGS observations of NLS1s 
are reported on Mrk~766 and Mrk~359.
Absorption features from \ion{O}{v-viii} are detected 
in Mrk~766 and is very similar to that of MCG$-$6-30-15
\citep{braRay01, sako03}.
On the other hand, Mrk~359 does not show absorption lines 
but rather emission lines from \ion{O}{vii}  and \ion{Ne}{ix}
\citep{obr01}.

It is still difficult to have unified view of warm absorber 
residing in NLS1s.
There are two objects lacking the detection of warm absorber:
Ton~S180 is the most luminous among the six objects, whereas
the other of Mrk~359 is less luminous than 
the Ark~564 and Mrk~766 with warm absorber.
Even though luminosity could be considered to be one of 
the primary parameters to make variety of warm absorber,
we cannot simply explain this difference 
only by the difference in luminosity.
Regarding the outflow, 
to launch it  by radiation in the gravity from a central black hole, 
higher luminosity normalized by Eddington ($L/L_{E}$)
is required (Reynolds \& Fabian 1995). 
However, NGC~4051 with the fastest outflow has the lowest $L/L_{E}$,
if we estimate the central mass from the reverberation mapping 
technique (Kaspi et al. 2000).
Again, the outflow velocity seems difficult to be explained 
simply by one parameter of $L/L_{E}$.

We also examined optical spectra from Ton~S180 and NGC~4051 (kindly
supplied by Jules Halpern), Ark~564 (from HST archive), and
Mrk~478 and Mrk~766, and MCG$-$6-30-15.  
We looked for differences in the [\ion{O}{iii}] lines from
the narrow-line region, the broad component of H$\beta$, as well as
\ion{He}{ii} and coronal lines  which have been suggested to be associated
with the warm absorber (Peterson et al. 2000; Porquet et al.\ 1999).
The clearest difference that we found was that Ton S180 and Mrk~478 
lacked coronal line emission (Here we looked at the \ion{Fe}{x} and \ion{Fe}{vii}
lines), 
while all of the other objects had
coronal lines.  Furthermore, Ton~S180 and Mrk~478 had 
blueward-asymmetric
[\ion{O}{iii}] lines, while the other objects had rather symmetric [\ion{O}{iii}]
lines.  Since Ton~S180 and Mrn~478 seem unusual in our comparison for its lack of
X-ray absorption lines and low equivalent width FUV absorption lines,
this result perhaps supports the idea that the warm absorber is
associated with coronal line emission.

\subsection{The 1~keV Emission Line-like Feature}

The \chandra\  observation allowed us to 
advance our understanding of the spectrum of Ark~564.
We found no prominent narrow emission line around 1~keV
and ruled out the possibility that the 1~keV feature
originates from blends of several narrow emission lines. 
However, the  possibility remains that 
a broad emission line could be present
that is not easily seen in the somewhat low-signal-to-noise
spectrum.

To investigate a broad emission line,
we fitted the MEG and HEG first order spectra binned at \asca\ 
resolution 
with a model consisting of a power-law, blackbody and a Gaussian line
near 1~keV. The line width $\sigma$ was fixed at 100~eV.
The fit gives a center energy of $1.02\pm0.03$~keV and 
the EW of $18\pm7$~eV. 
The energy is consistent with the results from \asca\ 
but the EW is smaller.

Although the detected absorption lines with the EWs of 
$\sim$1~eV cannot be seen in spectra with medium energy
resolution, the predicted absorption edges may be apparent. 
We speculate that the 1~keV feature is an artifact of the edges.
In order to investigate this point, 
we add both the inferred UTA and 
the edges from the low and high-$\xi$ 
warm absorbers, fixing the parameters to be best-fit values,
to the model. The spectral model {\sl absori} in {\sl XSPEC}
was used for 
the model of the edges of warm absorbers.
Then, the center energy of the broad line 
becomes  $1.02^{+0.05}_{-0.04}$~keV 
and the EW becomes  $9{\pm7}$~eV. 
Although the decrease of the EW is not statistically significant, 
the EW is smaller and nearly consistent with zero. 
Furthermore, the fit with the two warm absorbers
was not significantly improved 
by adding the broad line component to the model
(85~\% significance by F-test).
Thus the 1~keV feature appears to be consistent with being
an artifact of the warm absorber,
although we cannot completely rule out a broad emission line or
blends of many narrow absorption lines.

\section{Summary}

From our \chandra\  HETGS observation,
we confirmed the presence of the soft excess.
If we fitted the excess with blackbody model, 
the best-fit temperature is 0.124~keV.
No prominent narrow emission lines are observed around 1~keV.
We detected an edge-like feature at 0.712~keV in the 
source rest frame. 
The preferred interpretation of this feature is combination of 
the O~{\sc vii} K-edge and
a number of the L-absorption lines from the slightly ionized iron,
which suggests a warm absorber with $\xi\sim1$ 
and $N${\sc h}$~\sim10^{21}$~cm$^{-2}$\@.
These properties are roughly consistent with those of
the UV absorber.
We also detected narrow absorption lines of O~{\sc vii}, 
O~{\sc viii}, Ne~{\sc ix}, Ne~{\sc x}, and Mg~{\sc xi}
at the systemic velocity. From these lines, another 
warm absorber having  $\log\xi\sim$ 2 and 
$N${\sc h}$~\sim10^{21}$~cm$^{-2}$ is required.

\acknowledgments

We acknowledge the great efforts of 
all the members of the \chandra\  team.
We are grateful to Prof. Jules Halpern 
for reduction of optical spectra and helpful discussion.
We thank John Moore for his analyses of optical spectra.
We also thank an anonymous referee for careful reading
and helpful comments.
CM and KML gratefully acknowledge support 
through the \chandra\ GO 0-11624 grant.
HLM is supported by SAO contract SV1-61010 to MIT.

\clearpage

\clearpage 

\begin{figure}
\plotone{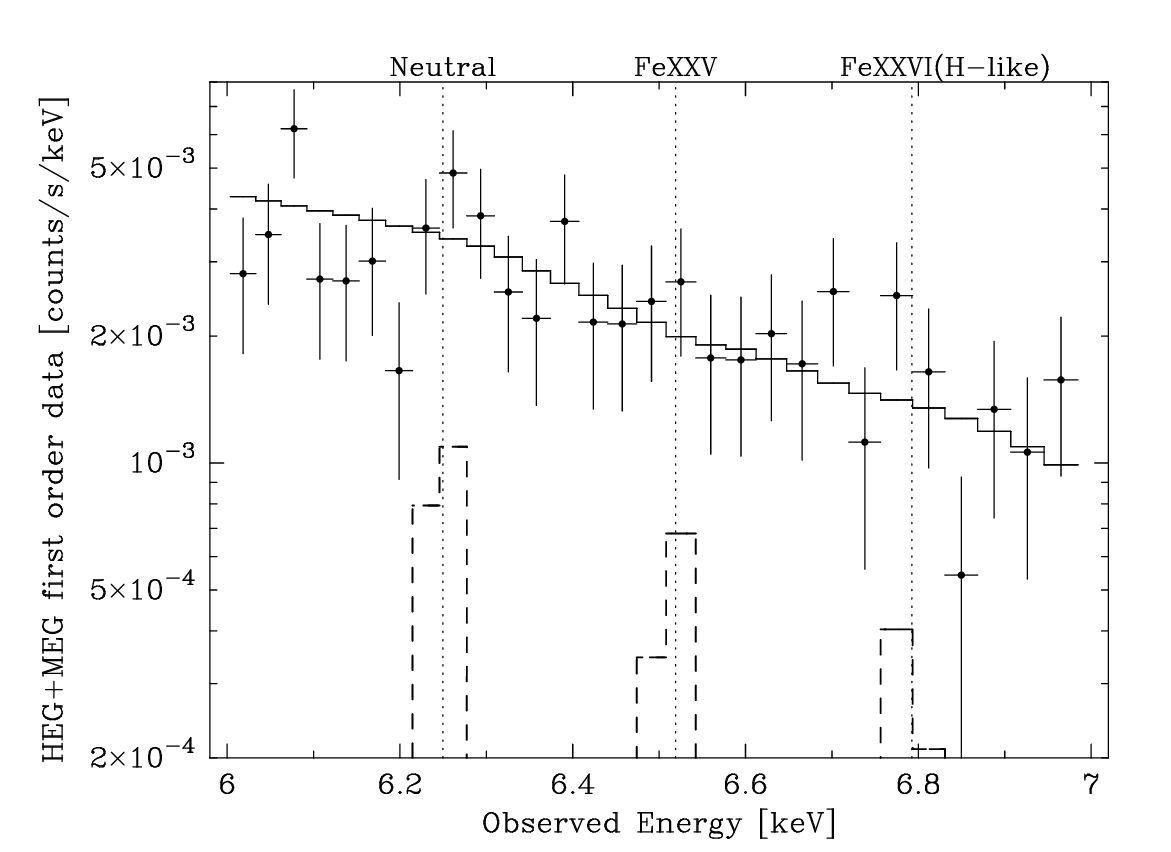}
\caption{The HETGS spectrum in the iron K-line energy band.
The HEG$\pm$1 and MEG$\pm$1 are added for plot.
The solid stepped line shows power-law continuum level.
The energies of neutral, He-like and H-like iron K-line are marked
with the dotted lines.
For reference, the lines with the EW of 20~eV 
are shown by the dashed line.
}
\label{fig:iron}
\end{figure}

\clearpage

\begin{deluxetable}{lcl}
\tablecaption{The 2--10~keV flux state during the \chandra\  and \asca\ observations.
The \chandra\ data are from the MEG and HEG $\pm1$ orders.
The background is negligible in HETGS.  The \asca\  data are from the GISs. 
The background is  subtracted.
 \label{tab:var}}
\tablewidth{0pt}
\tablehead{
\colhead{Observation} 
& \colhead{Count Rate [c/s]}%%[c s$^{-1}$]}   
& \colhead{RMS \tablenotemark{a}}  
}
\startdata
~\chandra         			&0.259$\pm$0.002& 24.6$\pm$1.6         \\
~\asca\ simultaneous with \chandra     	&0.659$\pm$0.007& 24.8$\pm$2.0 \\
~\asca\ whole (6/1--7/6)  		&0.546$\pm$0.001& 25.1\tablenotemark{b}
\enddata
\tablenotetext{a}{The time bin width is 256~s.}
\tablenotetext{b}{The RMS was calculated in every interval with the same 
span as the \chandra\  observation, then their average is calculated.
}
\end{deluxetable}

\clearpage

\begin{deluxetable}{lclllll}
\tabletypesize{\scriptsize}
\tablecaption{Parameters of narrow absorption lines 
from the Gaussian fit. 
All lines are unresolved.
\label{tab:line}}		
\tablehead{
\colhead{$E_{obs.}$\tablenotemark{a}} 
& \colhead{$E_{lab}$\tablenotemark{b}} 
& \colhead{EW}
& \colhead{Identification}
& \multicolumn{3}{c}{EWs of possible accompanying lines from  the ion\tablenotemark{c}} \\
\colhead{[eV]}	& \colhead{[eV]} & \colhead{[eV]} & \colhead{ion \& transition} & 
	\colhead{$\alpha$}	&\colhead{$\beta$}	&\colhead{$\gamma$}
}
\startdata
~654.0$^{+0.7}_{-1.3}$ & 653.5   & 0.68$^{+0.36}_{-0.32}$      &O~{\sc viii} Ly$\alpha$
        & ---       & 0.71$\pm0.33$    & $-0.33(<0.40)^f$\\
~666.2$^{+0.4}_{-0.9}$ & 665.6  & 0.74$^{+0.41}_{-0.29}$      &O~{\sc vii} 1s$^2$--1s3p 
        & 0.19($<0.69$) & ---       & $-$0.06($<0.32$)  \\
~721.1$^{+0.6}_{-0.5}$ & (719--722)	& 0.57$^{+0.29}_{-0.29}\,^{e}$ & (\ion{Fe}{v-vi})  
	&---	&---	&---	\\
~775.4$^{+0.4}_{-0.6}$\,$^d$ & 774.6   & 0.69$^{+0.25}_{-0.26}$      &O~{\sc viii} Ly$\beta$
        &0.75$\pm0.40$ & ---        & $-0.33(<0.40)^g$\\
~897.5$^{+0.5}_{-1.0}$  & (896.8) & 0.59$^{+0.32}_{-0.32}\,^f$ & (\ion{Fe}{xvii})
	&---	&---	&---	\\
~922.6$^{+0.7}_{-0.4}$\,$^d$ & 922.0  &0.92$^{+0.37}_{-0.43}$   &Ne~{\sc ix} 1s$^2$--1s2p 
        &---        &       0.76$\pm0.38$&  ~~0.26($<0.76$)  \\

1022.5$^{+0.8}_{-0.5}$\,$^d$ & 1021.9   & 1.84$^{+0.42}_{-0.38}$      &Ne~{\sc x} Ly$\alpha$(+ \ion{Fe}{xvii})
        & ---       & 0.13($<0.62$)  & ~~0.49($<1.09$)\\ 

1073.8$^{+0.6}_{-1.4}$ &1073.7  & 0.76$^{+0.57}_{-0.40}$      &Ne~{\sc ix} 1s$^2$--1s3p 
        & 0.79$\pm0.42$ & ---       & ~~0.26($<0.76$)  \\
1353$^{+1}_{-1}$ &1352.2  & 1.15$^{+0.58}_{-0.50}$ &Mg~{\sc xi} 1s$^2$--1s2p 
        &---        & 0.48($<1.29$)         & ~~0.43($<1.21$)  \\
\enddata
\tablenotetext{a}{The best-fit value of the line central energies in the source rest frame.}
\tablenotetext{b}{The theoretical line energies from references}
\tablenotetext{c}{The EWs are derived from the Gaussian fits at fixed energies
with fixed velocity width of 0, which result in slightly different values from 
the EWs in the third column 
from the Gaussian fits with free center energy and free velocity 
width.}
\tablenotetext{d}{The 90~\% lower limit is slightly larger than $E_{lab}$ 
by 0.1--0.2~eV (corresponding velocity of several ten~km~s$^{-1}$).
}
\tablenotetext{e}{ The EW is subject to the continuum modeling
	because it is part of a broad deficit. 
	The velocity width of $\sigma$ is fixed at 0.}
\tablenotetext{f}{The central energy is fixed at the best-fit value 
	during the EW error estimation. $\sigma=0$ is assumed.}
\tablenotetext{g}{Only the MEG$+$1 data are used, because this feature 
	in the MEG$-$1 data falls in the chip gap.}
\end{deluxetable}

\end{document}